\documentstyle[epsfig,12pt,twoside]{article}

\newcommand{\be}{\begin{equation}}
\newcommand{\ee}{\end{equation}}
\newcommand{\bea}{\begin{eqnarray}}
\newcommand{\eea}{\end{eqnarray}}
\newcommand{\bean}{\begin{eqnarray*}}
\newcommand{\eean}{\end{eqnarray*}}

\begin{document}



\title{New Study of the Isotensor  $\pi\pi$  Interaction
                }

\vspace{2cm}
\author{F.Q. Wu$^b$, B.S.Zou$^{a,b,c}$, L.Li$^{b,d}$, D.V.Bugg$^e$ \\
a) CCAST (World Laboratory), P.O. Box 8730, Beijing 100080\\
b) Institute of High Energy Physics, CAS, Beijing 100039, China
\footnote{mailing address, E-mail: zoubs@mail.ihep.ac.cn}\\
c) Institute of Theoretical Physics, CAS, Beijing 100080, China\\
d) Peking University, Beijing 100087, China\\
e) Queen Mary College, London, UK}

\date{\today}

\maketitle

\begin{abstract}
  With t-channel $\rho$, $f_2(1270)$ exchange and the $\pi\pi\to\rho\rho\to\pi\pi$ box
  diagram contribution, we reproduce the $\pi \pi$ isotensor S-wave and D-wave
  scattering phase shifts and inelasticities up to 2.2 GeV
  quite well in a K-matrix formalism. The t-channel $\rho$
  exchange provides repulsive negative phase shifts while the
  t-channel $f_2(1270)$ gives an attractive force to increase the phase shifts
  for $\pi \pi$ scattering above 1 GeV, and the coupled-channel box diagram
  causes the inelasticities. The implication to the isoscalar
  $\pi\pi$ S-wave interaction is discussed.
\end{abstract}
\vspace{1cm}

\newpage

\section{Introduction}

Much attention has been paid to the isospin I=0 $\pi\pi$ S-wave
interaction due to its direct relation to the $\sigma$ particle
and the scalar glueball candidates
\cite{PDG,Ochs,BSZ,Gasser,Anisovich,Achasov,Zou93}. However, to
really understand the isoscalar $\pi\pi$ S-wave interaction, one
must first understand the isospin I=2 $\pi\pi$ S-wave interaction
due to the following two reasons:  (1) There are no known
s-channel resonances and less coupled channels in I=2 $\pi\pi$
system, so it is much simpler than the I=0 $\pi\pi$ S-wave
interaction; (2) To extract I=0 $\pi\pi$ S-wave phase shifts from
experimental data on $\pi^+\pi^-\to\pi^+\pi^-$ and
$\pi^+\pi^-\to\pi^0\pi^0$ obtained by $\pi N\to\pi\pi N$
reactions, one needs an input of the I=2 $\pi\pi$ S-wave
interaction.

Experimental information on the I=2 $\pi\pi$ scattering mainly
came from $\pi^+p\to\pi^+\pi^+n$ \cite{Hoogland77} and
$\pi^-d\to\pi^-\pi^-pp$ \cite{Durusoy73,Cohen} reactions. As shown
in Fig.\ref{rho}, the main features for the I=2 $\pi\pi$ S-wave
phase shifts $\delta_0^2$ and inelasticities $\eta_0^2$ are: (1)
the $\delta_0^2$ goes down more and more negative as the $\pi\pi$
invariant mass increases from $\pi\pi$ threshold up to 1.1 GeV;
(2) the $\delta_0^2$ starts to increase for energies above about
1.1 GeV; (3) the $\eta_0^2$ starts to deviate from 1 for energies
above 1.1 GeV. The first feature can be well explained by the
t-channel $\rho$ exchange force
\cite{Isgur,lilong,Zou94,Speth,Locher,Oset} although it can also
be reproduced by other approaches \cite{Ishida}. Due to the
relative poor quality of the I=2 $\pi\pi$ scattering data above
1.1 GeV, the other two features are usually overlooked. In this
paper, we show in a K-matrix formalism \cite{lilong,lilong2} that
these two features can be well reproduced by the t-channel
$f_2(1270)$ exchange and the $\pi\pi$-$\rho\rho$ coupled-channel
effect, respectively.

A correct description of the $I=2$ $\pi\pi$ S-wave interaction is
important for the extraction of the $I=0$ $\pi\pi$ S-wave
interaction from experimental data. While the $I=2$ $\pi\pi$
S-wave interaction can be extracted from the pure $I=2$
$\pi^\pm\pi^\pm\to\pi^\pm\pi^\pm$ reactions, the $I=0$ $\pi\pi$
S-wave interaction can only be extracted from
$\pi^+\pi^-\to\pi^+\pi^-$ and $\pi^+\pi^-\to\pi^0\pi^0$ reactions
which are mixture of $I=0$ and $I=2$ contributions. Recently, the
$\pi^+ \pi^-\to \pi^+ \pi^- $ scattering from the old $\pi N$
scattering experiments with both unpolarized\cite{CM} and
polarized targets\cite{Becker} has been re-analyzed\cite{BSZ,
Kaminski} in combination with new information from $p\bar p$ and
other experiments. The $\pi^+\pi^-\to\pi^0\pi^0$ scattering has
also been studied by E852\cite{E852} and GAMS\cite{GAMS}
Collaborations. The relation between the
$\pi^+\pi^-\to\pi^+\pi^-$, $\pi^0\pi^0$ S-wave amplitudes and the
isospin decoupled amplitudes is as the following:

\bea T_{s}(+-,+-) &=& T_s^{I=0}/3+T_s^{I=2}/6 ,\label{tpm}\\
 T_{s}(+-,00) &=& T_s^{I=0}/3-T_s^{I=2}/3. \label{t00} \eea

Experimental information on $T_s^{I=0}$ was extracted from
$T_{s}(+-,+-)$ and $T_{s}(+-,00)$ information by assuming some
kind of $T_s^{I=2}$ amplitude. For example, in
Refs.\cite{BSZ,lilong,CM}, a scattering length formula for I=2
$\pi \pi$ S-wave was used; in Ref.\cite{Achasov} another empirical
parametrization was used. These parametrizations give similar
phase shifts up to 1.1 GeV, but differ at higher energies. All
these previous analyses have ignored the inelastic effects in the
$I=2$ channel. We will demonstrate that the correct description of
the $I=2$ S-wave interaction has significant impact on the
extraction of the $I=0$ $\pi\pi$ S-wave amplitude for energies
above 1.1 GeV.

\section{Formalism}

\ \ \ \ In our normalization the S-matrix of two-body scattering
takes the form \be <p_1^{\prime} p_2^{\prime}\mid S \mid p_1 p_2
> = I-(2\pi)^4
\delta^4(p_1+p_2-p_1^\prime-p_2^\prime)\frac{T(p_1,p_2,
p_1^\prime,p_2^\prime)}{(2E_1)^{\frac{1}{2}}(2E_2)^{\frac{1}{2}}
(2E_1^\prime)^{\frac{1}{2}}(2E_2^\prime)^{\frac{1}{2}}},\ee the
states are normalized as

\be <p^\prime \mid p>=(2\pi)^3\delta^3(p-p^\prime).  \ee

Our normalization is such that the unitarity relation for
partial-wave amplitudes reads: \be \mathrm{Im} T_l(s)=
T^\dagger_l(s)\rho(s)T_l(s),\ee with $\rho(s)$ the diagonal matrix
of phase space.  Taking the $\pi\pi$ channel as channel 1, then
$\rho_1(s)=(1-4m_\pi^2/s)^{1/2}$.  The partial-wave amplitudes are
obtained from the full amplitude by the standard projection
formula\cite{Martin1974,lilong}
\be T_l(s)=\frac{1}{s-4m_\pi^2}\int_{4m_\pi^2-s}^{0}d t P_l(x)
[1+\frac{2t}{s-4m_\pi^2}]T(s,t,u), \ee where $P_l(x)$ is Legendre
function, and $s, t, u$ are the usual Mandelstam variables. The
relation between the amplitude $T_l(s)$ for $\pi\pi\to\pi\pi$
scattering and phase shift parameters $\delta_l$ and $\eta_l$ is
\be T_l^I(s)=\frac{\eta_l^I(s)e^{2i\delta_l^I(s)}-1}{2i\rho_1(s)}.
\ee

To exhibit the isospin structure, T can  be written in terms of
three invariant amplitudes A, B and C by \be
T(s,t,u)=A(s,t,u)\delta_{\alpha\beta}\delta_{\gamma\delta}+
B(s,t,u)\delta_{\alpha\gamma}\delta_{\beta\delta}+
C(s,t,u)\delta_{\alpha\delta}\delta_{\beta\gamma},\ee where
$\alpha, \beta, \gamma$ and $\delta$ are isospin indices of pions.
Using isospin projection operator leads to: \bea
T^{I=0}(s,t,u) &=& 3A(s,t,u) + B(s,t,u)+C(s,t,u) \\
T^{I=1}(s,t,u) &=& B(s,t,u) - C(s,t,u)           \\
T^{I=2}(s,t,u) &=& B(s,t,u) + C(s,t,u). \eea

Then we follow the $K$-matrix formalism as in
Refs.\cite{lilong,lilong2,Locher}. The $K$-matrix unitarization is
introduced by \be T^I_l(s)=\frac{K^I_l(s)}{1-i\rho(s)K^I_l(s)},
\ee in which $l$ is the index of partial-wave and I the index of
isospin.

For the two-channel case, the two-dimensional $K$ matrix and
$\rho(s)$ matrix are \be K=\left (
\begin{array}{ccc}
K_{11}  &  K_{12}  \\
K_{12}  &  K_{22}
\end{array} \right ),\hspace{1cm}
\rho(s)=\left (
\begin{array}{ccc}
\rho_1(s)  &  0  \\
0  &  \rho_2(s)
\end{array} \right ),
\ee where $\rho_2(s)$ is the phase space factor for $\rho\rho$
channel, $K_{11}$ is K-matrix element for the channel of $\pi\pi
\rightarrow \pi\pi$,
 $K_{22}$ for the channel of $\rho\rho \rightarrow \rho \rho
$, and $K_{12}$ for $\pi\pi \rightarrow \rho \rho$. Then
substituting relation Eq.(13) into $ T={K}/(1-i\rho K)$, one
obtains \be
T_{11}=\frac{K_{11}-i\rho_2(K_{11}K_{22}-K_{12}K_{21})}{1-i\rho_1
K_{11}-i\rho_2 K_{22} -\rho_1\rho_2(K_{11}K_{22}-K_{12}K_{21})},
\ee Ignoring the interaction between $\rho\rho$, we have
$K_{22}=0$; then
\be T_{11}=\frac{K_{11}+i K_{12} \rho_2 K_{21}}{1-i
\rho_1(K_{11}+iK_{12} \rho_2 K_{21} )}, \ee
where $K_{12} \rho_2 K_{21}$ corresponds to the contribution of
the $\pi\pi\to\rho\rho\to\pi\pi$ box diagram. After substituting
$T_{11}$ for $T_l^I$ in Eq.(7), we may get phase shift parameters
$\delta$ and $\eta$ by solving Eq.(7).

In order to obtain $K_{11}$,  we incorporate the $t$-channel
$f_2(1270)$ contribution into the $t$-channel $\rho$ exchange term
by the Dalitz-Tuan method \cite{BSZ,lilong}. Supposing two
components $a$ and $b$ for the partial wave $l$ are expressed
individually as
$$T^a_l(s)=\frac{G_{a}}{C_a(s)-iG_{a}\rho_1(s)} \textrm{   and   }
T^b_l(s)=\frac{G_{b}}{C_b(s)-iG_{b}\rho_1(s)}, \nonumber$$ then
the combined amplitude will be \bea
\hat{T}^{ab}_l(s)&=&\frac{G_{a}C_b(s)+G_{b}C_a(s)}{[C_a(s)-iG_{a}\rho_1(s)]
[C_b(s)-iG_{b}\rho_1(s)]}  \nonumber  \\
&=&\frac{G_{ab}}{C_{ab}(s)-iG_{ab}\rho_1(s)}, \eea with
$C_{ab}(s)=C_a(s)C_b(s)-G_{a}G_{b}\rho^2_1(s)$, and
$G_{ab}=G_{a}C_b(s)+G_{b}C_a(s)$. The amplitude is explicitly
unitary, and the denominator contains the same poles as in each
component. Further poles are added one by one using the equations
above.
For example, the combination of the t-channel $\rho$ exchange and
the $f_2$ exchange gives \be
K_{11}=\frac{K_{\rho}(s)+K_{f_2}(s)}{1-\rho_1^2(s)K_{\rho}(s)
K_{f_2}(s)}.\ee

\subsection{$t$-channel $\rho$ meson exchange amplitude}

\ \ \ \ This section has  been partially presented in
Ref.\cite{lilong}. We start with the Born term of the $\pi\pi$
scattering amplitude by $\rho$ exchange \bea
T^{Born}(I=0)&=&2G_{\pi\pi\rho}[(\frac{\Lambda^2-m_\rho^2}{\Lambda^2-t})^2
           \frac{s-u}{m_\rho^2-t}+(\frac{\Lambda^2-m_\rho^2}{\Lambda^2-u})^2
           \frac{s-t}{m_\rho^2-u}],\\
T^{Born}(I=2)&=&-{1\over 2}T^{Born}(I=0), \eea where
$G_{\pi\pi\rho}=g^2_{\pi\pi\rho}/32\pi$=0.364, and at each vertex
we have used a form factor of conventional monopole type to take
into account the off-shell behavior of the exchanged mesons: \be
F(q^2)=\frac{\Lambda^2-m^2}{\Lambda^2-q^2}, \ee where m and q are
the mass and four-vector momentum, respectively, of exchanged
mesons, and $\Lambda$ is the cutoff parameter to be determined by
experimental data.

Their S-wave projections are \bea
K^{I\!=\!0}_S(s)&\!=\!&4G_{\pi\pi\rho}\{(\frac{m^2_\rho}{\Lambda^2}-1)
\frac{\Lambda^2+2s-4m^2_\pi}{\Lambda^2+s-4m^2_\pi}+ \nonumber
\\&&
\frac{2s+m^2_\rho-4m^2_{\pi}}{s-4m^2_{\pi}}ln{\frac{(s+m^2_\rho-4m^2_\pi)
\Lambda^2}{(s+\Lambda^2-4m^2_\pi)m^2_\rho}}\},\nonumber\\
& & \\
K^{I\!=\!2}_S(s)&\!=\!&-{1\over 2}K^{I=0}_S(s). \eea

Their I=2 D-wave projection is
\begin{eqnarray}
 K_{D}^{I=2}(s) &=&
 \frac{2G_{\pi\pi\rho} ( {\Lambda}^{2}-m_{\rho}^{2} )}
  {{\Lambda}^{2} (s-4m_{\pi}^{2})^{2}({\Lambda}^{2}+s-4m_{\pi}^{2})}
  \Big\{ 12{\Lambda}^{6} - 6{\Lambda}^{4} \big( 12m_{\pi}^{2} +
  m_{\rho}^{2} -4s \big)
  \nonumber \\ & & \hspace*{-8mm}
  + {\Lambda}^{2} \big( 28m_{\pi}^{2} + 6m_{\rho}^{2} -13s \big) \big( 4m_{\pi}^{2}-s \big)
  -2 \big( 2m_{\pi}^{2}-s \big) \big( s- 4 m_{\pi}^{2} \big)^{2}
  \Big\} \nonumber \\ & & \hspace*{-8mm} +
  \frac{G_{\pi\pi\rho}}{(4m_{\pi}^{2}-s)^{2}} \Big\{ 2
  \Big[ 12{\Lambda}^{6} + 12{\Lambda}^{2} m_{\rho}^{2}
  ( 8m_{\pi}^{2}-3s ) - 6{\Lambda}^{4} (
  8m_{\pi}^{2}+3m_{\rho}^{2} \nonumber \\  & & \hspace*{-8mm}
  -3s )
   + (4m_{\pi}^{2}-s) \big[ 16m_{\pi}^{4}
  +s(13m_{\rho}^{2}+2s) -4m_{\pi}^{2}(7m_{\rho}^{2}+3s) \big] \Big]
  {\ln} ( 1+  \nonumber \\ & & \hspace*{-8mm}
  \frac{s-4m_{\pi}^{2}}{{\Lambda}^{2}} )
  - 2(4m_{\pi}^{2}-m_{\rho}^{2}-2s) \Big[
  16m_{\pi}^{4} + 6m_{\rho}^{4}+6m_{\rho}^{2}s +s^{2}
  \nonumber \\ & &  \hspace*{-8mm}
  -8m_{\pi}^{2}(3m_{\rho}^{2}+s) \Big] {\ln}
  (1+\frac{s-4m_{\pi}^{2}}{m_{\rho}^{2}} ) \Big\}.
 \end{eqnarray}

\subsection{$t$-channel $f_2(1270)$ meson exchange amplitude }

\ \ \ \ The amplitude for the t-channel $f_2(1270)$ meson exchange
without considering the vertex form factor was given in
Ref.\cite{Zou94} for studying $I=0$ $\pi\pi$ scattering. Here we
include the form factor to take into account the off-shell
behavior of the exchanged $f_2$ meson. Then the Born amplitudes
for the $t$-channel $f_2$ exchange are \bea T^{Born}(I=0) &=&
G_{\pi\pi f_2
}[\frac{3(t-u)^2-(4m^2_\pi-s)^2}{m^2_{f_2}-s}(\frac{\Lambda^2+{m_{f_2}}^2}{\Lambda^2+s})^2+
\nonumber  \\ &&
\frac{(s-u)^2-(4m_\pi^2-t)^2/3}{m^2_{f_2}-t}(\frac{\Lambda^2-{m_{f_2}}^2}{\Lambda^2-t})^2+
\nonumber  \\ &&
\frac{(s-t)^2-(4m_\pi^2-u)^2/3}{m^2_{f_2}-u}(\frac{\Lambda^2-{m_{f_2}}^2}{\Lambda^2-u})^2],  \\
     T^{Born}(I=2) &=&
G_{\pi\pi f_2
}[\frac{(s-u)^2-(4m_\pi^2-t)^2/3}{m_{f_2}^2-t}(\frac{\Lambda^2-{m_{f_2}}^2}{\Lambda^2-t})^2+
\nonumber \\&&
\frac{(s-t)^2-(4m_\pi^2-u)^2/3}{m^2_{f_2}-u}(\frac{\Lambda^2-{m_{f_2}}^2}{\Lambda^2-u})^2],
\eea with $ G_{\pi\pi f_2} \simeq 0.19 \  GeV^{-2}$  determined
from the width of $f_2(1270)$.

Their S-wave projections are \bea K_{f_2}^{I=0} &=& K_{f_2}^{I=2}
\nonumber  \\ &=& G_{\pi\pi f_2}\{ \frac{\frac{4}{3}(m^2_{f_2}-\Lambda^2)(\Lambda^4+
6\Lambda^2s-8\Lambda^2m^2_\pi+6s^2-24sm^2_\pi+16m^4_\pi)}{\Lambda^2(\Lambda^2+s-4m^2_\pi)}
 \nonumber \\&&+\frac{2[(2s+m^2_{f_2}-4m^2_\pi)^2-(m^2_{f_2}-4m^2_\pi)^2/3]}{s-4m^2_\pi}
\ln [1+\frac{s-4m^2_\pi}{m^2_{f_2}}]  \nonumber \\ &&
+\frac{\frac{4}{3}(\Lambda^4-2\Lambda^2m^2_{f_2}-6s^2+24sm^2_\pi-6sm^2_{f_2}
-16m^4_\pi+8m^2_{f_2}m^2_\pi)}{s-4m^2_\pi}\times  \nonumber \\ &&
\ln [1+\frac{s-4m^2_\pi}{\Lambda^2}]\}. \eea

Their I=2 D-wave projection is

\begin{eqnarray}
 K_{D}^{I=2}(s)\!\!\!\!&=&
\frac{2G_{\pi\pi f_2}} {3\Lambda^2{( 4{m_{\pi}}^2 - s )
        }^3( \Lambda^2 - 4{m_{\pi}}^2 + s
      ) } \Bigg\{ -2
       ( \Lambda^2 - m_{f_2}^2 )
         ( 4{m_{\pi}}^2 -  \nonumber \\ &&
         s )
       \Big[ 18\Lambda^8 -
3\Lambda^6( 60{m_{\pi}}^2 +
            2{m_{f_2}}^2 - 31s )  +
         2{( -4{m_{\pi}}^2 + s )
              }^2( 8{m_{\pi}}^4  \nonumber \\ && -
            12{m_{\pi}}^2s + 3s^2 )-
          \Lambda^2( 4{m_{\pi}}^2 - s
            )\Big( 128{m_{\pi}}^4 -
            6{m_{f_2}}^4 +    \nonumber \\ &&
            4{m_{\pi}}^2
             ( 15{m_{f_2}}^2 - 44s
               )  - 39{m_{f_2}}^2s +
            42s^2 \Big)  +
         \Lambda^4\Big( 544{m_{\pi}}^4 -  \nonumber \\ &&
            6{m_{f_2}}^4 +
            28{m_{\pi}}^2
             ( 3{m_{f_2}}^2 - 20s
               )  - 45{m_{f_2}}^2s +
            112s^2 \Big)  \Big]  +
      2\Lambda^2( \Lambda^2 -    \nonumber \\ &&
      4{m_{\pi}}^2 +
         s )\Bigg[ \Big( 18\Lambda^8 -
            12\Lambda^6( 12{m_{\pi}}^2 +
               2{m_{f_2}}^2 - 7s )  -   \nonumber \\ &&
            2\Lambda^2{m_{f_2}}^2
             ( 304{m_{\pi}}^4 -
               344{m_{\pi}}^2s + 73s^2
               )  -
            2( 4{m_{\pi}}^2 - s )
               ( 32{m_{\pi}}^6 -     \nonumber \\ &&
               3s^2
                ( 7{m_{f_2}}^2 + s )
                   + 8{m_{\pi}}^2s
                ( 11{m_{f_2}}^2 +
                  3s )
             {} - 8{m_{\pi}}^4
                ( 8{m_{f_2}}^2 + 7s
                  )  )  +     \nonumber \\ &&
            \Lambda^4 ( 304{m_{\pi}}^4 +
               8{m_{\pi}}^2
                ( 27{m_{f_2}}^2 -
                  43s )  +
               s( -126{m_{f_2}}^2 +    \nonumber \\ &&
                  73s ) ) \Big)
          \log [\frac{\Lambda^2}{\Lambda^2-4{m_{\pi}}^2 - s}] +
         \Big( {( -4{m_{\pi}}^2 +
                {m_{f_2}}^2 ) }^2 +
            6( -4{m_{\pi}}^2 +      \nonumber \\ &&
               {m_{f_2}}^2 ) s + 6s^2
            \Big)
            \Big ( 16{m_{\pi}}^4 +
            6{m_{f_2}}^4 +
            6{m_{f_2}}^2s  + s^2 -
            8{m_{\pi}}^2
             ( 3{m_{f_2}}^2 +    \nonumber \\ && s )
            \Big) \log [\frac{{m_{f_2}}^2}{{m_{f_2}}^2-4{m_{\pi}}^2 + s}]
            \Bigg]\Bigg\} .
\end{eqnarray}

\subsection{The $\pi\pi\to\rho\rho\to\pi\pi$ box diagram amplitude}

\begin{figure}[htbp] \vspace{-0.6cm}
\begin{center}
\includegraphics[scale=1.2]{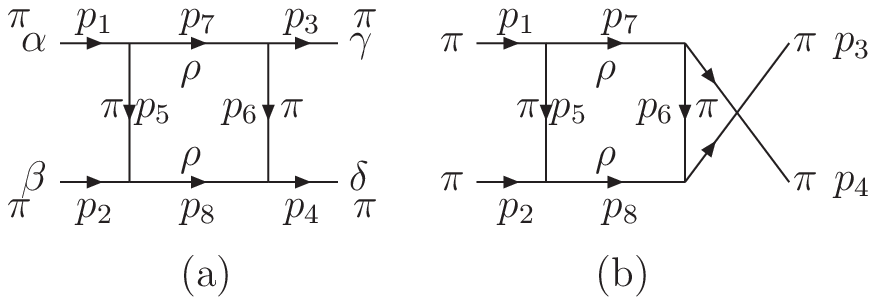}
 \caption{The $\pi\pi\to\rho\rho\to\pi\pi$ box diagrams}
 \label{block1}
\end{center}
\end{figure}

For energies above the $\rho\rho$ threshold, the inelastic effect
should be taken into account in the $I=2$ $\pi\pi$ channel. Note
that unlike $I=0$ $\pi\pi$ channel, the $I=2$ $\pi\pi$ channel
does not couple to the $K\bar K$ channel due to isospin
conservation. The $\pi\pi$ channel couples to the $\rho\rho$
channel by the t-channel $\pi$ exchange as shown in
Fig.\ref{block1}.

\ \ \ \ \ Assuming the on-shell approximation \cite{Lu} for the
$\rho\rho$ intermediate state, the Lorentz-invariant matrix
element $\mathcal{M}$ of the box diagram (Fig.\ref{block1}(a)) is
 \bea
  i{\mathcal{M}}_a &=& {}-\frac{1}{64\pi^2} \cdot \rho_2 \cdot
{g_{{\pi}{\pi}{\rho}}^4} \int d\Omega_7 \Big[-(p_1+p_5) \cdot
(p_3-p_6)+ \nonumber \\ &&
 \frac{(p_1+p_5) \cdot p_7 \cdot p_7 \cdot
(p_3-p_6)}{{m_\rho}^2}\Big]^2   \cdot
\frac{1}{(p_5^2-m_\pi^2)(p_6^2-m_\pi^2)}\cdot  \nonumber \\ &&
(\delta_{\alpha\beta}\delta_{\gamma\delta}+\delta_{\alpha\gamma}\delta_{\beta\delta}),
\label{amplitude} \eea
where we have used the Cutkosky rule \cite{field}.
$\alpha,\beta,\gamma $ and $ \delta$ are their isospin indices.
The box diagram contribution includes Fig.\ref{block1} (a) and
Fig.\ref{block1} (b). Fig.\ref{block1} (b) gives the same
amplitude as Fig.\ref{block1} (a), so we have
${\mathcal{M}}={\mathcal{M}}_a+{\mathcal{M}}_b=2{\mathcal{M}}_a$.
In our convention, $ T={\mathcal{M}}/(32\pi) $. After including
the vertex form factors to take into account the off-shell
behavior of the exchanged pions, its $I=2$ S-wave projection is
 \bea
 T_s^{I=2} &=& \frac{1}{2}\int d \cos{\theta_3}T \nonumber \\&=&
i\cdot \frac{G_{{\pi}{\pi}{\rho}}^2}{2\pi} \cdot \rho_2 \cdot
\int d\cos\theta_3 d\cos\theta_7 d\phi_7 \Big[-(p_1+p_5) \cdot
(p_3-p_6)+ \nonumber
\\&&
 \frac{(p_1+p_5) \cdot p_7 \cdot
p_7 \cdot (p_3-p_6)}{{m_\rho}^2}\Big]^2  \cdot
\frac{1}{(p_5^2-m_\pi^2)(p_6^2-m_\pi^2)}   \nonumber\\
& & \cdot
(\frac{\Lambda_{\pi\pi\rho}^2-m_\pi^2}{\Lambda_{\pi\pi\rho}^2-p_5^2})^2
\cdot
(\frac{\Lambda_{\pi\pi\rho}^2-m_\pi^2}{\Lambda_{\pi\pi\rho}^2-p_6^2})^2
 \eea
with $G_{{\pi}{\pi}{\rho}} = g_{{\pi}{\pi}{\rho}}^2/(32\pi)=0.364
$. This box diagram amplitude is related to the K-matrix element
$K_{12}$ as

\be T_s^{I=2} = i K_{12} \rho_2 K_{12} .\ee

The $\rho_2$ is the phase space for $\rho \rho$. If we ignore the
width of the $\rho$ meson to treat it as a stable particle, the
$\rho_2$ in our convention can be written as
   \bea
    \rho_2^{stable}(s) &=& (8\pi)\cdot (2\pi)^4 \int \delta^4 (p-p_{\rho_1}-p_{\rho_2})
    \frac{d^3 p_{\rho_1}}{(2\pi)^32 E_{\rho_1}}
    \frac{d^3 p_{\rho_2}}{(2\pi)^32 E_{\rho_2}} \nonumber \\
    &=& \frac{2{\mid p_{\rho}\mid}}{\sqrt{s}}=\sqrt{1-{4M^2_\rho\over s}}.
   \label{rho2s}
   \eea
Taking into account the width of the $\rho$ meson, $\Gamma_\rho$,
then the phase space factor should be
   \bea
    \rho_2(s) &=& \frac{1}{\pi^2} \int d s_{12} d s_{34} \frac{M_\rho
\Gamma_\rho}{(M_\rho^2-s_{12})^2+(M_\rho
    \Gamma_\rho)^2} \cdot  \frac{M_\rho \Gamma_\rho}{(M_\rho^2-s_{34})^2+(M_\rho
\Gamma_\rho)^2} \nonumber  \\ &&
\cdot\frac{\sqrt{[s-(\sqrt{s_{12}}+\sqrt{s_{34}})^2][s-(\sqrt{s_{12}}-\sqrt{s_{34}})^2]}}{s}.
   \label{rho2}
   \eea

\begin{figure}[htbp]
\vspace{-1.6cm}
\begin{center}
\includegraphics[width=10cm,height=10cm]{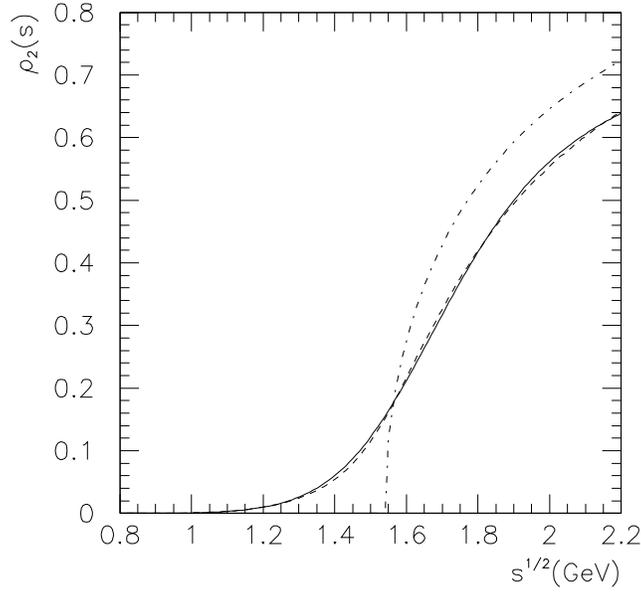}
\caption{Comparison of $\rho\rho$ phase space $\rho_2(s)$ given by
Eq.(\ref{rho2s}) (dot-dashed line), Eq.(\ref{rho2}) (dashed line)
and Eq.(\ref{rho2a}) (solid line). } \label{fig:rho2}
\end{center}
\end{figure}

   In order to get analytic formula of $\rho_{2}(s)$ for convenience of
   application, we use the following formula to fit $\rho_{2}(s)$
   obtained numerically by Eq.(32)
   \bea
    \rho_{2}(s)= \frac{\sqrt{1-\frac{1}{s}}}{1+e^{a_1s^3+a_2s^2+a_3 s
    +a_4}} \label{rho2a}
   \eea where $a_1$, $a_2$,  $a_3 $ and $a_4$ are parameters with fitted values
   as $a_1=-0.0888, a_2=1.3337, a_3=-6.9465, a_4=11.5204$.
   This analytic formula for $\rho_2(s)$ is used in our final
   calculation. Fig.\ref{fig:rho2} shows the comparison of this analytic formula
   (solid line) with those given by Eq.(\ref{rho2s}) (dot-dashed
   line) and Eq.(\ref{rho2a}) (dashed line).

\begin{figure}[htbp]
\vspace{-1.6cm}
\begin{center}
\includegraphics[width=12cm,height=16cm]{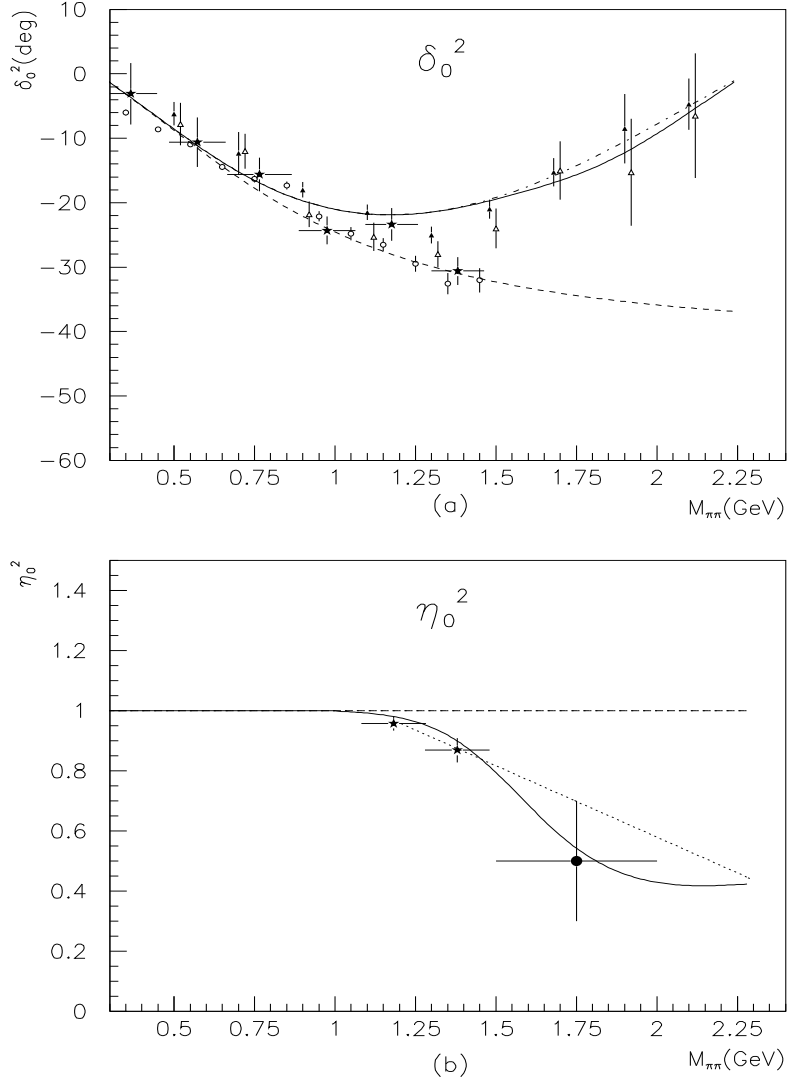}
\caption{The $I=2$ $\pi\pi$ $S$-wave phase shift $\delta_0^2$ (a)
and inelastic coefficient $\eta_0^2$ (b). The experimental data
for $\delta_0^2$ are from  Ref.\cite{Hoogland77} (circles),
Ref.\cite{Cohen}(stars), and Ref.\cite{Durusoy73}(triangles); the
data for $\eta_0^2$ are from Ref.\cite{Cohen}(stars), and
Ref.\cite{Durusoy73} (solid circle). The solid curves are results
including the total contribution from $\rho$, $f_2(1270)$ exchange
and the box diagram. The dot-dashed curves are from $\rho$ and
$f_2(1270)$ exchange. The dashed curves include only t-channel
$\rho$ exchange. The dotted line in (b) is
$\eta_0^2=1.53-0.475m_{\pi\pi} (GeV/c^2)$ used in
Ref.\cite{Durusoy73}. } \label{rho}
\end{center}
\end{figure}

\begin{figure}[htbp]
\vspace{-1.6cm}
\begin{center}
\includegraphics[width=12cm,height=16cm]{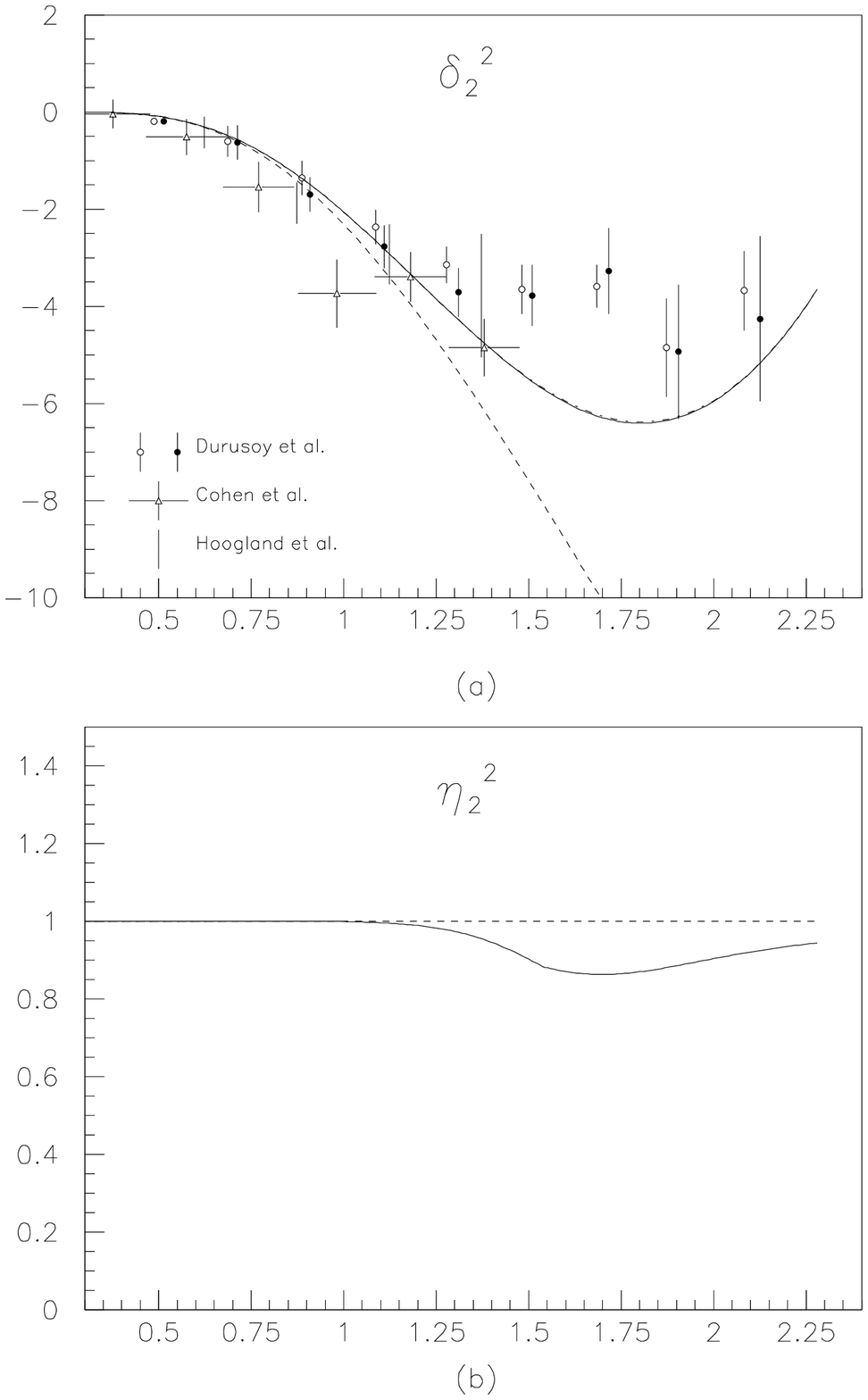}
\caption{The $I=2$ $\pi\pi$ $D$-wave phase shift $\delta_2^2$ (a)
and inelastic coefficient $\eta_2^2$ (b). The experimental data
for $\delta_2^2$ are from Ref.\cite{Hoogland77} (short lines),
Ref.\cite{Durusoy73}(squares), and Ref.\cite{Cohen}(triangles).
The solid curves represent the total contribution of $\rho$,
$f_2(1270)$ exchange and  the box diagram. The dot-dashed curves
are from $\rho $ and $f_2(1270)$ exchange. The dashed curves
include only t-channel $\rho$ exchange.} \label{phase}
\end{center}
\end{figure}

\section{Numerical results and discussion}

\ \ \ \ From the formalism given above, we get I=2 $\pi\pi$ S-wave
and D-wave phase shifts and inelasticities as shown in
Fig.\ref{rho} and Fig.\ref{phase}.  The t-channel $\rho$ exchange
alone (dashed lines) reproduces the phase shifts for energies up
to 1.1 GeV very well with the form factor parameter
$\Lambda_{\rho\pi\pi}=1.5$ GeV \cite{lilong}, but underestimates
the phase shifts at higher energies. The inclusion of the
t-channel $f_2(1270)$ exchange increases the phase shifts
especially for energies above 1 GeV and can reproduce the phase
shift data very well with the form factor parameter
$\Lambda_{f_2\pi\pi}=1.7$ GeV as shown by the dot-dashed lines.
However the t-channel $\rho$ and $f_2$ exchange only contribute to
the elastic scattering and cannot produce the inelasticities for
energies above 1 GeV.

The experimental information on the inelasticities is scarce for
the $I=2$ $\pi\pi$ scattering. Two data points were given by
Ref.\cite{Cohen} for energies $1\sim 1.5$ GeV. For energies
$1.5\sim 2$ GeV, Ref.\cite{Durusoy73} estimated to be $0.5\pm 0.2$
for the $\eta^2_0$ in one solution and assumed
$\eta_0^2=1.53-0.475m_{\pi\pi}$ (GeV/$c^2$) for another solution.
The two solutions gave similar results for the $I=2$ $\pi\pi$
S-wave phase shifts. For the $I=2$ $\pi\pi$ D-wave scattering, the
inelasticity could not be measured well and was ignored, {\sl
i.e.}, assuming $\eta^2_2=1$ for the extraction of the
$\delta^2_2$ shown in Fig.\ref{phase}.

\begin{figure}[htbp]
\vspace{-0.5cm}
\begin{center}
\includegraphics[scale=1.1]{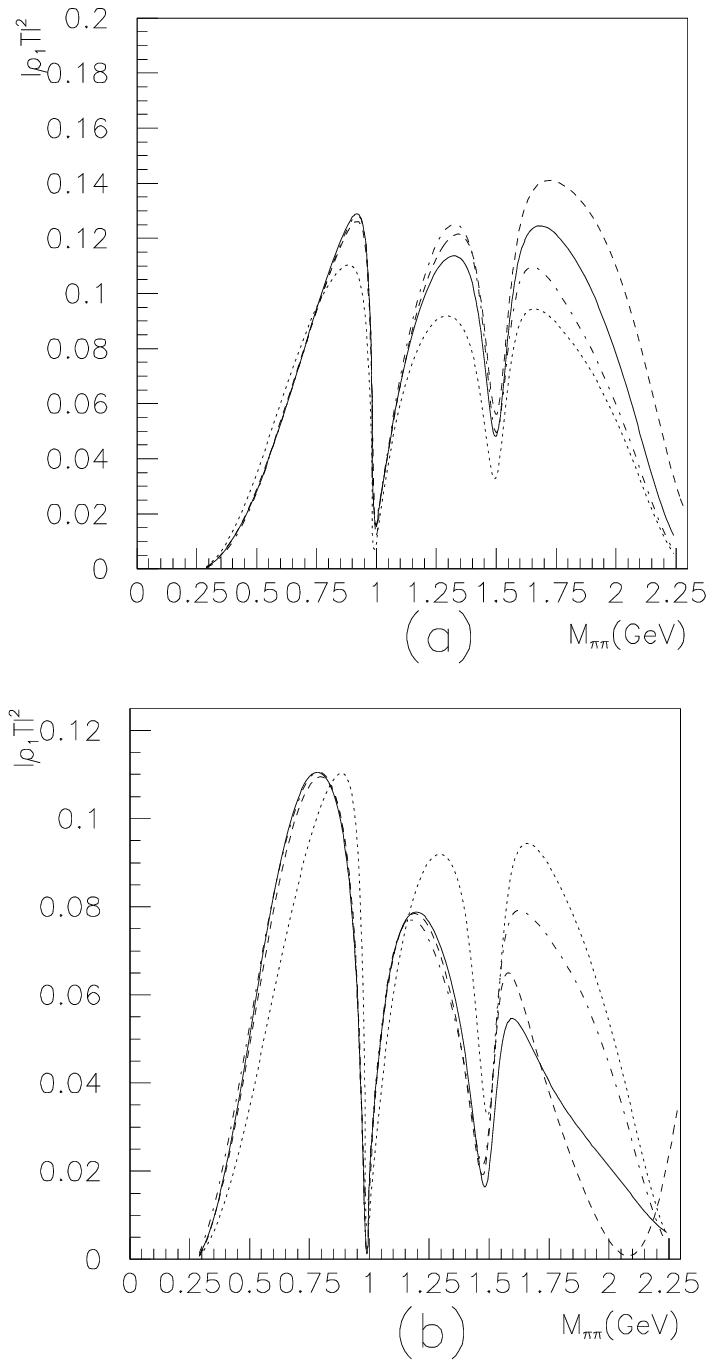}
\caption{Full amplitudes squared of $\pi^+\pi^-\to\pi^+\pi^- $ (a)
and $\pi^+\pi^-\to\pi^0\pi^0$ (b). The lines correspond to using
$T^{I=0}_s$ from Ref.\cite{lilong} plus various input of
$T^{I=2}_s$: $T^{I=2}_s=0$ (dotted lines); $T^{I=2}_s$ from
Refs.\cite{lilong,CM} (dot-dashed lines); $T^{I=2}_s$ from
Ref.\cite{Achasov} (dashed lines); $T^{I=2}_s$ from this work
including t-channel $\rho$ \& $f_2$ exchange and the
$\pi\pi\leftrightarrow\rho\rho$ coupled-channel effect. }
 \label{sampl}
\end{center}
\end{figure}

Although there are only three data points with large error bars
for the inelasticity parameter $\eta^2_0$ of the $I=2$ $\pi\pi$
S-wave scattering, it is clear that the inelastic effect may be
significant around 1.6 GeV. In order to reproduce this
inelasticity, it is necessary to consider the
$\pi\pi\leftrightarrow\rho\rho$ coupling channel effect. We find
that including contribution from the $\pi\pi\to\rho\rho\to\pi\pi$
box diagram in our K-matrix formalism, the $I=2$ S-wave
inelasticity data can be very well reproduced without introducing
any more free parameter as shown in Fig.\ref{rho}(b). The same
diagram also predicts a broad shallow dip around 1.7 GeV for the
inelasticity of the $I=2$ D-wave scattering as shown in
Fig.\ref{phase}(b). Assuming $\eta^2_2=1$ for energies around 1.7
GeV may bias the $\delta^2_2$ data around this energy. This may be
the reason that the $\delta^2_2$ data around 1.7 GeV has the
largest discrepancy with our theoretical result. The box diagram
has little influence to the phase shifts although it produces
large inelasticity.

Inspired by the new discovery of a pentaquark state
\cite{Nakano,Russia,CLAS,SAPHIR}, we also explored the possibility
of including an $I=2$ s-channel resonance to reproduce the $I=2$
$\pi\pi$ S-wave scattering data instead of using the t-channel
$f_2$ exchange and the $\pi\pi\to\rho\rho\to\pi\pi$ box diagram,
but failed to reproduce the $\delta^2_0$ and $\eta^2_0$ data
simultaneously. While the $\eta^2_0$ needs the resonance with mass
around 1.6GeV, the $\delta^2_0$ needs the mass above 2.3 GeV with
a much broader width.

To demonstrate the significance of possible impact of the $I=2$
input for the extraction of $I=0$ $\pi\pi$ amplitude, we calculate
the full S-wave amplitudes for  $\pi^+ \pi^- \to \pi^+ \pi^- $ and
$\pi^+ \pi^- \to \pi^0 \pi^0 $ according to
Eqs.(\ref{tpm},\ref{t00}) with $T^{I=0}_s$ from Ref.\cite{lilong}
plus various $T^{I=2}_s$ inputs. The corresponding full S-wave
amplitudes squared ($|\rho_1T|^2$) are shown in Fig.\ref{sampl}.
The dotted lines are the results with $T^{I=2}_s=0$. The
dot-dashed lines correspond to the scattering length formula for
I=2 S-wave as $T_s^{I=2}=a_0 q /(1-i a_0 q)$ and
$a_0=-0.11\pm0.01m_{\pi}^{-1} $ as used in
Refs.\cite{BSZ,lilong,CM}, which is similar to the result by
considering only the t-channel $\rho$ exchange contribution. The
dashed lines use the new empirical $T^{I=2}_{s}$ formula of
Ref.\cite{Achasov}, which is similar to the result by considering
t-channel $\rho$ and $f_2$ exchange contributions, but ignoring
the inelasticity caused by $\pi\pi\to\rho\rho\to\pi\pi$ box
diagram contribution. The solid lines are results with our
$T^{I=2}_{s}$ including the t-channel $\rho, f_2$ exchange and the
contribution of the box diagram. It is clear that $T^{I=2}_s$ has
significant contribution to the amplitudes for
$\pi^+\pi^-\to\pi^+\pi^-$ and $\pi^+\pi^-\to\pi^0\pi^0$ processes,
hence has significant impact on the extraction of the $T^{I=0}_s$
amplitude. Previous inputs of $T^{I=2}_s$ give similar results as
our new $T^{I=2}_s$ for energies below 1.1 GeV, but differ from
ours significantly for higher energies. The inclusion of both
t-channel $f_2$ exchange and $\pi\pi\to\rho\rho\to\pi\pi$ box
diagram contribution is important.

In summary, the basic features of $I=2$ $\pi\pi$ scattering phase
shifts and inelasticities can be well reproduced by the t-channel
meson ($\rho$,$f_2$) exchange and the
$\pi\pi\leftrightarrow\rho\rho$ coupled-channel effect in the
K-matrix formalism. The t-channel $\rho$
  exchange provides repulsive negative phase shifts while the
  t-channel $f_2(1270)$ gives an attractive force to increase the phase shifts
  for $\pi \pi$ scattering above 1 GeV, and the coupled-channel box diagram
  causes the inelasticities. A correct description of the $I=2$ $\pi\pi$
scattering has significant impact on the extraction of the $I=0$
scattering amplitudes from $\pi^+\pi^-\to\pi^+\pi^-$ and
$\pi^+\pi^-\to\pi^0\pi^0$ data, especially for energies above 1.2
GeV . A re-analysis of these data with our new description of the
$I=2$ $\pi\pi$ scattering will be carried out as our next step.

\section{Acknowledgments} The work is partly supported by CAS
Knowledge Innovation Project (KJCX2-SW-N02) and the National
Natural Science Foundation of China under Grant
Nos.10225525,10055003. We thank the Royal Society for funds
allowing a collaboration between Queen Mary College, University of
London and IHEP, Beijing.

\end{document}